# Free Agent in Agent-Based Mixture-of-Experts Generative AI Framework

Jung-Hua Liu
National Chung Cheng University
emileliu@ccu.edu.tw

## 1. Introduction

Multi-agent systems have long been central to research in artificial intelligence (AI), aiming to emulate or extend human-like collaboration and problem-solving in distributed environments. Recent breakthroughs in Generative AI (Gen AI)—particularly with Large Language Models (LLMs) and advanced deep learning techniques—have regained interest in multi-agent frameworks. Although large-scale generative models now excel at tasks involving text, images, and other modalities, integrating them into multi-agent (or agent-based) systems poses both opportunities and challenges.

In current multi-agent Gen AI implementations, each agent typically specializes in a specific role: one might handle text summarization, another code generation, yet another data analysis, and so forth. While this specialized, role-based approach capitalizes on targeted expertise, it can become rigid and not generalizable. Agents are usually designed, trained, and assigned to fixed tasks, and there is little or no mechanism to dynamically swap out underperforming agents or reassign expertise. Yet real-world teams—whether in business, sports, or professional ecosystems—often replace or add participants to adapt to evolving conditions and maintain a competitive edge.

Drawing from Major League Baseball (MLB), we introduce the "free agent" concept. In professional sports, free agents allow teams to replace players who are underperforming or no longer fit the team's needs. This continual infusion of new talent encourages ongoing performance improvements and resilient strategies. Translating this to AI, this paper proposes the **Reinforcement Learning Free Agent (RLFA)** algorithm, which formalizes the notion of a free agent in multi-agent Gen AI. Through a straightforward reward mechanism, poorly performing agents can be removed while higher-performing replacements seamlessly take their place.

A key enhancement in this paper is the use of a **mixture-of-experts (MoE)** approach within the multi-agent framework. MoE is a deep learning architecture in which multiple sub-models—or "experts"—each address specific aspects of a task, while a gating function assigns inputs to the most relevant expert. By integrating multi-agent methods with MoE, each agent—whether long-standing or newly introduced as a free agent—can harness specialized sub-models to optimize performance for distinct data modalities or problem domains.

In practical scenarios such as financial reporting, medical diagnosis, or fraud detection, multi-agent systems distribute tasks among themselves, calling upon specialized modules as needed. When an agent consistently underperforms or fails to meet certain metrics, the system identifies a suitable free agent to replace it. This free agent, also guided by reinforcement learning, is rewarded not only for producing accurate results (e.g., in fraud detection) but also for maintaining strong synergy with other agents, respecting privacy guidelines, and making efficient use of resources. This paper shows that the free agent mechanism can significantly boost both performance and adaptability in multi-agent Gen AI systems. It also provides a straightforward route for continuous upgrades—new, more capable models can join as free agents, incrementally displacing those that are no longer effective. This is particularly important for critical tasks, such as fraud detection, which must not be hindered by outdated or poorly performing agents.

## 2. Literature Review

### 2.1. Multi-Agent Systems and Their Increasing Popularity

Advances in large language models and generative AI have spurred many organizations to investigate multi-agent frameworks for more scalable and intricate applications (Cruz, 2024; Rasheed et al., 2024). In essence, multi-agent systems break tasks into sub-tasks, each handled by autonomous agents capable of local decision-making and inter-agent communication. This division enhances concurrency, modularity, and parallel processing, which is invaluable for complex tasks like data analysis, knowledge retrieval, summarization, and content generation.

Microsoft's **Magentic-One** exemplifies a multi-agent platform that incorporates specialized modules (e.g., for natural language processing or code debugging), all orchestrated by an overarching controller (Fourney et al., 2024). Similarly, OpenAI's **Swarm** (Bigio, 2024) organizes workflow segments—termed "routines and handoffs"—to enable parallelization and avoid overloading any single specialized agent. While these systems showcase the benefits of multi-agent orchestration, they typically assume permanent agent roles, lacking an automated mechanism for replacing persistently underperforming agents.

Amazon has proposed **Knowledge Graph Enhanced Language Agents (KGLA)** for recommendation systems, which deploy separate agents to simulate user behavior, identify correct purchase intents, and track incorrect intents (Guo et al., 2024). Li et al. (2024) further noted that sparse agent communication topologies might improve collective reasoning by allowing more time for consensus-building. However, an automated approach for removing or upgrading ineffective agents has remained largely unexplored.

### 2.2. Mixture-of-Experts (MoE) in Generative AI

Parallel to multi-agent research, the **mixture-of-experts (MoE)** strategy has become pivotal in deep learning. MoE entails multiple "expert" networks, each fine-tuned for a specialized aspect or subset of the overall task domain, and a "gating" network that selects which expert(s) to utilize for each input. This technique excels in heterogeneous environments or tasks requiring diverse skill sets.

Zhang et al. (2024) illustrated how combining multi-agent coordination with MoE yields high-fidelity 3D object generation in VR/AR, with distinct sub-experts handling textures, geometry, and lighting. MoE has also been applied to real-time, data-intensive domains, including intelligent transportation systems, where a gating function routes traffic or incident data to experts trained on those specific phenomena (Xu et al., 2024). By merging agent-based task orchestration with MoE-level specialization, systems achieve more robust, context-sensitive performance.

**2.3. Existence of Incompetent Agents**

A significant concern in multi-agent systems is the persistence of underperforming or "incompetent" agents. Model drift, limited training data, or new domain requirements can erode an agent's effectiveness over time (Motwani et al., 2024). Existing multi-agent frameworks typically rely on human administrators to identify and replace stagnant agents, rather than automatically employ an internal mechanism for competitive improvement. Addressing this gap, the **RLFA** algorithm introduces a market-like dynamic where agents can be automatically swapped, akin to professional sports free agency.

**2.4. Fraud Detection and Free Agents**

Fraud detection exemplifies why an automated system for replacing outdated agents is critical. Tasks commonly involve parsing communications, detecting anomalies, and monitoring user behaviors. If any of these specialized agents fail to adapt to evolving tactics—such as new phishing methods—the entire operation's security is jeopardized. By adopting a free agent concept akin to MLB, the system promptly dismisses underperforming agents and adds newly retrained or specialized models that address emerging fraud vectors. A reinforcement learning reward schema guides the new agent to optimize fraud detection rates, minimize false positives, and preserve synergy with other agents.

## 3. Concept

### 3.1. Conceptual Overview of RLFA

The **Reinforcement Learning Free Agent (RLFA)** algorithm introduces a sports-inspired mechanism for replacing underperforming agents with stronger candidates. Drawing on Major League Baseball (MLB) free-agency rules, we propose that an agent reaches "free-agent status" after accruing a certain "service time" in the system or upon being "released" for subpar performance. Once an agent gains free-agency, it is no longer bounded by its initial "contract" and is able to be recruited by any "team" (i.e., any segment of the multi-agent framework needing a replacement).

**Service Time and Control**

In MLB, players become free agents after six years of Major League service or if they are released sooner. Analogously, our RLFA system measures an agent's "service time" using metrics such as completed tasks, successful episodes, or overall runtime in the environment. When an agent meets its required service time—and provided it continues to perform adequately—it remains under the system's control. Once that service time expires, the agent achieves free-agency and can "sign" with another subsystem or replace an incumbent agent, if doing so aligns with the system's strategic or performance goals.

**Release and Eligibility**

In MLB, a player released before reaching six years of service may sign a new contract yet does not have full free-agency rights until meeting the required threshold. By analogy, if an RLFA agent underperforms, it can be "released" before it fulfills its service time. Upon release, the agent enters the free-agent "pool," making it available for re-hiring—either by the same subsystem if its performance improves or by a different subsystem relying on its expertise. Until the agent accrues the necessary service time, however, it may face constraints such as limited data access or narrower task assignments, akin to MLB arbitration.

These free-agency constructs endow RLFA with flexibility and an inherent mechanism for continuous improvement. Underperforming agents do not linger indefinitely, while high-potential agents can join or rejoin at opportune moments.

**3.1.1. Reward Design**

In the Reinforcement Learning Free Agent (RLFA) approach, each agent is trained and evaluated using a reward mechanism that reflects performance across multiple dimensions. Building on traditional reinforcement learning paradigms, our methodology defines a **multi-factor reward** function for each agent $i$ at time $t$. Specifically,

$$R_i(t) = \alpha * \text{Accuracy}_i(t) + \beta * \text{Synergy}_i(t) + \gamma * \text{Efficiency}_i(t) - \delta * \text{Penalty}_i(t),$$

where:

1) $\text{Accuracy}_i(t)$: A normalized performance metric, such as classification accuracy, F1 score, or any task-specific measure of correctness.
2) $\text{Synergy}_i(t)$: A measure of how well the agent collaborates with other agents, reflecting communication quality, the frequency of successful task handoffs, or other indicators of multi-agent coordination.
3) $\text{Efficiency}_i(t)$: A ratio or score indicating the agent's resource consumption (e.g., processing time, memory usage) relative to a target baseline or reference policy.
4) $\text{Penalty}_i(t)$: A factor capturing negative outcomes such as misclassifications, policy violations, or excessive resource usage that surpasses acceptable thresholds.

The **weights** $\alpha, \beta, \gamma$, and $\delta$ are hyperparameters selected based on the importance of each component for the overall system objectives. For example, an application focusing on fraud detection with high stakes for errors may prioritize $\alpha$ (accuracy) more heavily, while a

collaborative environment might stress $\beta$ (synergy). The RLFA training loop is listed in the below:

1) **Initialization:**
   a) Each agent begins with random or pretrained parameters.
   b) A set of free agents is also initialized and remains on standby in a "free-agent pool."
2) **Task Assignment and Data Flow:**
   a) The system assigns relevant tasks (e.g., fraud detection, code generation) to each agent based on its role.
   b) Agents process input data, optionally leveraging internal mixture-of-experts (MoE) modules to select specialized sub-models.
3) **Reward Computation:**
   a) At the end of each episode or upon completing a designated task batch, each agent's performance metrics are evaluated.
   b) The multi-factor reward $R_i(t)$ is calculated using the formula above.
4) **Parameter Update:**
   a) Agents apply a reinforcement learning algorithm (e.g., Q-Learning, policy gradient, or actor-critic) to update their parameters based on $R_i(t)$.
   b) Higher rewards reinforce agent strategies that yield better accuracy, synergy, or efficiency.
5) **Agent Replacement:**
   a) At scheduled intervals or upon detecting poor performance (e.g., $R_i(t)$ below a threshold $\tau$ for a sustained period), the system "releases" underperforming agents into the free-agent pool.
   b) A new or retrained agent is introduced from the free-agent pool in a probationary ("shadow") mode. If it demonstrates sufficiently high $R_i(t)$, it fully replaces the incumbent.
6) **Convergence and Continuous Improvement:**
   a) Over multiple training cycles, underperforming agents are systematically replaced by those with higher rewards.
   b) The mixture-of-experts component within each agent can also adapt, refining internal sub-models that contribute to different aspects of $R_i(t)$.

**Practical Considerations**

- **Hyperparameter Tuning:** Determining optimal values for $\alpha, \beta, \gamma,$ and $\delta$ typically involves trial and error, guided by domain-specific expertise (e.g., placing more emphasis on accuracy in medical diagnostics).
- **Scalability:** As the number of agents or tasks increases, computing each agent's reward can become more

resource-intensive. Distributed computing methods may be necessary to maintain real-time or near-real-time performance.
- **Privacy and Security:** For sensitive domains like fraud detection, the RLFA architecture should incorporate privacy-preserving measures (e.g., anonymized data during the probationary phase) to mitigate potential risks of sharing information with newly introduced agents.

**3.1.2. Replacement Mechanics and Agent Integration**

Adopting MLB roster management principles, RLFA enforces a clear pathway for replacing agents:

1) **Performance Evaluation**
   At regular intervals, the system assesses each agent's metrics (e.g., accuracy, throughput, or synergy).
2) **Trigger Condition**
   If an agent's performance falls below a threshold for a designated period $\Delta t$, or if the agent is ill-suited to evolving tasks, it is "released" into the free-agent pool before consuming all its service time.
3) **Free-Agent Pool**
   Released agents, along with newly introduced models, reside in this pool. Their eligibility depends on performance expectations, synergy requirements, and resource constraints—akin to how MLB teams sign new players based on specific roles.
4) **Agent Signing and Integration**
   A free agent is signed if it delivers clear advantages, such as better performance or specialized skill sets. Initially, the new agent may operate in a probationary ("shadow") mode to verify compatibility.
5) **Probation to Full Integration**
   Should the free agent meet or surpass performance goals in its probation phase, it replaces the incumbent agent entirely and begins accruing its own service time.

**3.2. Mixture-of-Experts Integration**

While RLFA controls agent-level transitions, each agent can internally adopt a **mixture-of-experts (MoE)** paradigm. As in baseball, where specialized coaches focus on hitting, pitching, or fielding, each sub-expert within MoE handles a different dimension of an agent's functionality:

- **Expert Specialization**
  For example, a fraud-detection agent—akin to a "pitcher"—may have separate sub-experts specializing in text analytics, numeric analysis, or user-behavior profiles.
- **Gating Mechanism**
  Like a coaching staff choosing which pitch to throw, a gating function decides which sub-expert is most relevant for the input at hand.

Upon entering free agency, an agent retains or refines its MoE architecture to remain competitive in the free-agent pool. Consequently, RLFA and MoE work together to ensure each agent calls on the most appropriate sub-models, whether newly introduced or returning from the pool.

### 3.3. Fraud Detection Use Case

The fraud detection scenario clearly demonstrates how free agency accelerates adaptation in dynamic, adversarial domains:

1) **Data Ingestion**
   Continuous logs or messages arrive in the system, akin to a baseball season with daily games.
2) **Agent Dispatch**
   A "manager" agent assigns a fraud detection "pitcher" based on its skill in identifying emerging attack patterns.
3) **Sub-Expert Query**
   Leveraging MoE, the agent's specialized "coaches" scrutinize text, numeric data, and user behavior for anomalies.
4) **Scoring**
   Much like a pitcher's earned run average (ERA) or strikeout rate, an agent's metrics (e.g., true positives, precision, and recall) are tallied and tracked.
5) **RLFA Trigger**
   If an agent's accuracy consistently dips below α\alphaα, it is "cut" from the lineup (released) into the free-agent pool.
6) **Free-Agent Signing**
   A new or recently enhanced model, proven effective against current fraud tactics, can replace the incumbent after a probationary phase to confirm its performance.
7) In time, the system builds a dynamic roster of top-performing agents, each utilizing MoE sub-models for specialized tasks, as illustrated by the pseudocode in **Figure 1**, **Figure 2**, **Figure 3**, and **Figure 4**.

```
Algorithm 1: Fraud Detection Pipeline
Result: Fraud classification results for each data sample
Inputs:
    • IncomingDataQueue: Stream or batch of transactions/messages
    • SelectFraudAgent: Method to select the best FraudDetection agent
    • Agents: Current set of active agents (some specialized in fraud detection)
foreach dataSample in IncomingDataQueue do
    assignedAgent ← SelectFraudAgent(Agents);
    decision ← assignedAgent.detectFraud(dataSample);
    LogDetection(dataSample, decision);
    if decision == fraud then
        TakeAction(dataSample);
        // For instance: block the transaction, raise an
            alert, etc.
    end
    UpdateFraudMetrics(assignedAgent, dataSample, decision);
end
```

Figure 1. Pseudocode for Fraud Detection Pipeline

```
Algorithm 2: RLFA: Evaluate and Release Underperforming Agents
Result: Agents that underperform or exceed service time are released into the
        FreeAgentPool
Inputs:
    • Agents: Current active agents (including FraudDetection)
    • FreeAgentPool: Pool of non-active agents
    • alpha: Performance threshold (e.g., F1 score ¿= 0.80)
    • maxServiceTime: Maximum service time before automatic free agency
while system or environment is running do
    foreach agent in Agents do
        EvaluateFraudPerformance(agent);
        if (agent.performance ¡ alpha) OR (agent.serviceTime ¿= maxServiceTime) then
            releaseAgent(agent);
            Remove agent from Agents;
            Add agent to FreeAgentPool;
        end
    end
end
```

Figure 2. Evaluate and Release Underperforming Agents

```
Algorithm 3: RLFA: Filling Vacant Roles & Managing Probationary Agents
Result: Vacant roles are filled; probationary agents transition or are released
Inputs:
    • Agents: Active agents (some may be probationary)
    • FreeAgentPool: Pool of released or new candidate agents
    • alpha: Performance threshold
    • VacantRoles: Roles that must be filled
while system or environment is running do
    // --- Step 1:  Fill Vacant Roles ---
    foreach role in VacantRoles do
        requiredSkills ← getRequiredSkillsForRole(role);
        CandidateList ← filterCandidates(FreeAgentPool,
         requiredSkills);
        if CandidateList is not empty then
            BestCandidate ← selectAgent(CandidateList);
            Remove BestCandidate from FreeAgentPool;
            signAgent(BestCandidate, role);
            Add BestCandidate to Agents;
        end
    end
    // --- Step 2:  Transition Probationary Agents ---
    foreach agent in Agents do
        if agent.status == "Probationary" then
            agent_moe_init(agent);
            EvaluateFraudPerformance(agent);
            if agent.performance ¿= alpha then
                Promote agent to FullMember;
                agent.operational_mode ← "Active";
            else
                releaseAgent(agent);
                Remove agent from Agents;
                Add agent to FreeAgentPool;
            end
        end
    end
    // --- Step 3:  Increment Service Time ---
    foreach agent in Agents do
        agent.serviceTime ← agent.serviceTime + 1;
    end
end
```

Figure 3. Filling Vacant Roles & Managing Probationary Agents

```
Algorithm 4: EvaluateFraudPerformance
Result: Updates an agent's overall fraud detection performance metric
Inputs:
    • agent: The agent whose performance we are evaluating
    • PerformanceLog: Data structure storing aggregated results (true positives, false
      positives, false negatives) from UpdateFraudMetrics
TP ← PerformanceLog[agent].truePositives;
FP ← PerformanceLog[agent].falsePositives;
FN ← PerformanceLog[agent].falseNegatives;
Compute Precision:
                    Precision = TP / (TP + FP)
Compute Recall:
                    Recall = TP / (TP + FN)
Compute F1 Score:
                    F1 = 2 × (Precision × Recall) / (Precision + Recall)
if Precision + Recall ¿ 0 then
    agent.performance ← F1;
end
else
    agent.performance ← 0;
end
// Optionally compute additional metrics, e.g., AUC,
    specificity, etc.
```

Figure 4. EvaluateFraudPerformance

## 4. Analysis

This section discusses the conceptual impact of RLFA in agent-based systems with MoE. It addresses overall performance improvements, partial observability, and privacy/security factors, with a particular focus on fraud detection. While the discussion here is theoretical, actual implementations would require thorough experimental evaluations using metrics such as accuracy, F1 scores, or recall rates.

### 4.1. Performance in Task Completion

Multi-agent Gen AI systems often aim to increase both the volume and the quality of completed tasks. By employing RLFA:

- **Higher Overall Accuracy**
  Automated replacement prevents persistent performance deficits.
- **Rapid Adaptation**
  As tasks evolve—from text analytics to code generation—the system can introduce specialized free agents, reducing retraining overhead.
- **Competition-Driven Enhancement**
  A pool of free agents, each incentivized to surpass incumbents, spurs continuous improvements.

### 4.2. Partial Observability and Free Agent Unlocking Mechanism

Real-world tasks commonly involve partial observability, where agents do not share a complete global view. RLFA accommodates this via a progressive "unlocking" approach:

- **Restricted Access**
  During a free agent's probation, it sees only anonymized or partial data.

- **Expanded Permissions**
  As the free agent proves reliable, it gains access to more detailed or sensitive information.

This ensures new agents can neither compromise privacy nor degrade system performance before demonstrating competence.

**4.3. Privacy and Security Considerations**

In multi-agent systems handling sensitive data, any new agent introduces potential privacy risks. RLFA mitigates these by:

- **Strict Access Control**
  Free agents begin with limited data privileges.
- **Sandbox Testing**
  Unverified agents are tested in isolated environments with obfuscated data.
- **Performance-Driven Access**
  Free agents only receive expanded data access upon meeting trust and accuracy benchmarks.

Such staged integration is particularly vital in fraud detection, where a maliciously introduced agent could otherwise exploit data access immediately.

**4.4. Fraud Detection Analysis**

To illustrate RLFA's effectiveness:

- The incumbent fraud detection agent's accuracy drops from 95% to 75% with new scam patterns.
- RLFA identifies a better-trained or more robust agent from its pool.
- This free agent operates in shadow mode, where it achieves 88% before surpassing 90% in regular deployment.
- Once confirmed, it permanently replaces the incumbent, restoring or exceeding previous performance benchmarks.

Over multiple iterations, RLFA ensures timely updates and continuous fraud-detection enhancements, guarding against emergent threats.

# 5. Conclusion

This paper introduces the **Reinforcement Learning Free Agent (RLFA)** algorithm as a method to address two critical challenges in multi-agent Gen AI: (1) preventing the stagnation and suboptimal performance of incumbent agents, and (2) supporting specialized tasks that demand dynamic reconfiguration of agent skill sets. Drawing inspiration from Major League Baseball (MLB), RLFA adapts free agency to AI by replacing underperforming agents with better candidates, guided by a structured reward mechanism. This paper additionally demonstrates how a **mixture-of-experts (MoE)** architecture can strengthen the underlying agent capabilities, enabling each agent—permanent or free—to tap into specialized sub-models.

By combining multi-agent coordination with MoE, RLFA leads to both broader system coverage and deeper expertise.

**5.1. Key Contributions**

- **Adaptive and Competitive Environment**
  RLFA enforces a market-like dynamic, ensuring that poorly performing agents are replaced through transparent, reward-driven processes.
- **Enhanced Fraud Detection**
  Fraud detection benefits from ongoing adaptation, with newly introduced agents staying current on emerging malicious tactics.
- **Privacy and Security Management**
  Partial observability and restricted data access allow free agents to be incrementally integrated without compromising sensitive information.
- **Seamless MoE Integration**
  Specialized sub-models within each agent further amplify the advantages of dynamic agent replacement.

**5.2. Limitations and Future Directions**

- **Implementation Complexity**
  Orchestrating RLFA involves overhead in scheduling, evaluation, and gating.
- **Resource Constraints**
  Large-scale models can be expensive to deploy and maintain, warranting budget-aware RLFA variants.
- **Fairness and Bias**
  Ensuring free agents are unbiased requires careful training-data governance.
- **Real-World Validation**
  Comprehensive tests in production environments—particularly those subject to adversarial attacks and large-scale concurrency—are needed.

Possible future efforts include federated or decentralized implementations, where free agents could be shared across multiple organizations or devices with privacy safeguards. Another intriguing avenue is pairing RLFA with auto-curriculum learning, continuously generating tasks that challenge existing agents and facilitate further innovation.

**5.3. Final Remarks**

By adapting a free-agent paradigm from professional sports and integrating it with a mixture-of-experts framework, the **Reinforcement Learning Free Agent (RLFA)** approach promotes continuous improvement in multi-agent Generative AI systems. It offers a robust mechanism for removing stagnating or obsolete agents and seamlessly incorporating high-performing ones. As generative AI expands into domains ranging from healthcare to autonomous systems, the ability to dynamically upgrade agent-based architectures will be essential. RLFA, in tandem with MoE, can provide a promising pathway toward resilient, high-performance, and secure multi-agent AI ecosystems.